\documentclass[11pt,twoside]{article}

\usepackage{asp2006}
\usepackage{epsf}
\usepackage{psfig}
\usepackage{lscape}

\begin{document}
\title{Photospheric Manifestations of Supergranules During the Last Two Solar Minima}
\author{Peter E. Williams and W. Dean Pesnell}
\affil{Code 671, NASA Goddard Space Flight Center, Greenbelt, MD,
USA}

\begin{abstract}
Solar supergranulation plays an important role in generating and
structuring the solar magnetic field and as a mechanism responsible
for the 11-year solar cycle. It is clearly detected within SOHO/MDI
Dopplergrams, from which a variety of properties may be derived.
Techniques that extract spatial, temporal and kinematic
characteristics and provide comparisons for the two most recent
solar minima are described. Although supergranule lifetimes are
comparable between these minima, their sizes maybe slightly smaller
during the recent minimum.
\end{abstract}

\section{Introduction} \label{sec:Intro}

Supergranules are the largest scale component of solar convection
readily seen with present observing techniques. They are typically
$\sim$30 Mm across and have been observed to live from anywhere
between 1--2 days \citep{duv80}. Their relationship with local
magnetic fields have been seen via diffusion studies and spatial
observations of both magnetograms and CaIIK chromospheric data.

Since its inception, SOHO/MDI \citep{sch95} Dynamics Runs have
provided yearly 60-day sets of 1-minute Dopplergrams. Analysis of
the 1996 data has quantified a wide variety of supergranule
characteristics \citep[for example,][]{hat00, bec00}.

The most recent solar minimum has sparked much interest due to its
lengthy minimum. Many studies are currently underway to understand
the causes and consequences of this recent minimum.

This paper extends the analysis of supergranule characteristics to
2008 data and the results are compared to similar studies performed
on 1996 data.

\section{Methods} \label{sec:Methods}
A series of data reduction processes were applied to the
Dopplergrams to isolate the Doppler velocity signals of
supergranules. The analysis was then split into studies of
supergranule velocities, cell sizes and their 1/$e$ lifetimes.

\subsection{Data Reduction} \label{sec:Methods:DataRed}
The p-mode oscillations were removed by performing a weighted
average over 31 1-minute de-rotated Dopplergrams \citep{hat88}. The
resulting time-series was sampled every 15 minutes, producing 96
images per day. Next, the global flow fields were removed
\citep{hat87,hat92}. The remaining Doppler signals are due to the
photospheric convective velocity field. These convective
Dopplergrams were remapped to heliographic coordinates followed by a
projection onto the spherical harmonics to produce their respective
power spectra \citep{hat00}. Instrumental calibration artifacts were
also removed.

\subsection{Supergranule Velocities} \label{sec:Methods:Velocities}
\citet{hat02} performed a study of supergranule velocity flows using
the 1996 data and compared the relative strength of the radial and
horizontal components. That study was extended to investigate both
the 1996 and 2008 data, detailed by Williams \& Pesnell
\citetext{2010, in preparation}.

\subsection{Supergranule Lifetimes} \label{sec:Methods:Lifetimes}
\begin{figure}[!h]
\begin{center}
\plotfiddle{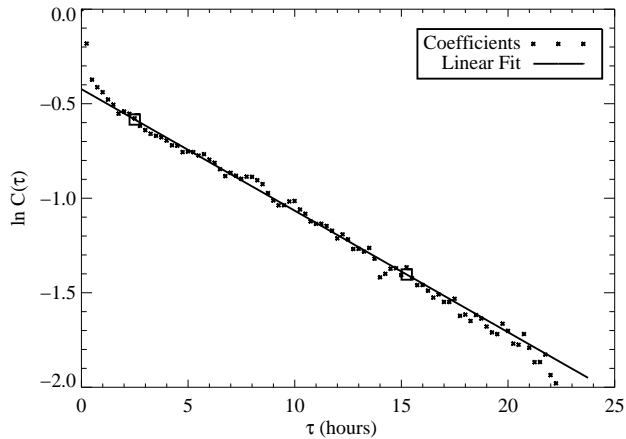}{2.1in}{0.0}{50.0}{50.0}{-130}{0}
\caption{A log-linear plot ({\itshape crosses}) of correlation
coefficient vs. time-lag represents the change of a supergranule
pattern during a single day. A linear fit to the plot ({\itshape
solid line}) extracts the decay coefficient for the convection
pattern and, subsequently, its 1/$e$ decay-time. The squares signify
the points between which the fit was performed. For this particular
day (in this case May 4, 2008), the decay coefficient is 0.0643,
giving a decay-time of 16 hours.} \label{fig:lifetime}
\end{center}
\end{figure}

Correlation techniques have been used to follow supergranule
patterns in the Dopplergrams as they traverse across the solar disk
and measure the change in the pattern over the period of a day. A
data strip, extracted near the equator, was selected from an image.
A similar strip window in a subsequent image, at a given time-lag
from the original, was moved longitudinally prograde across the
image and the correlation noted for each pixel shift. This process
was performed for all images within a single day, resulting in a
two-dimensional array (pixel-shift vs. time-lag) of correlation
coefficients. The region of decaying positive correlation pertaining
to the evolving supergranule pattern was isolated using a Radon
transform, from which the maximum correlation coefficient,
$C(\tau)$, at a given time lag, $\tau$, was found. A plot of $\ln
C(\tau)$ vs. $\tau$ was linearly fitted, the slope of which provided
the decay coefficient of the correlation (Figure
\ref{fig:lifetime}). The inverse of this decay coefficient gave the
1/$e$ decay-time of the supergranule pattern.

Performing this process for the 1996 \& 2008 data resulted in a
series of decay-times for each day. Treating each decay-time as an
independent result, a harmonic mean provided the average decay-time
for each data-set.

\subsection{Supergranule Sizes} \label{sec:Methods:Labels}
\begin{figure}[!h]
\begin{center}
\plotfiddle{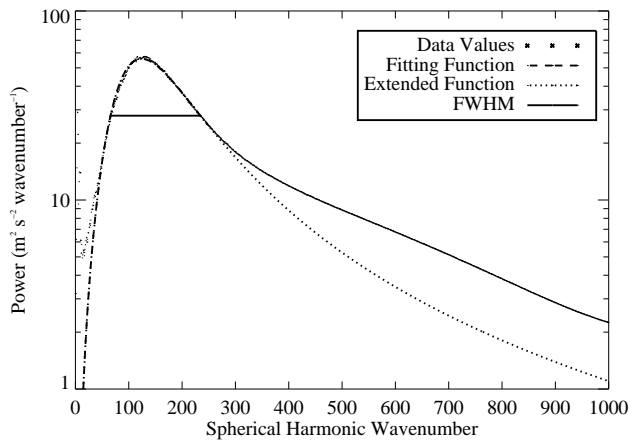}{2.1in}{0.0}{50.0}{50.0}{-130}{0}
\caption{A 60-day average of MDI convection spectra from 2008
({\itshape points}) is fit with a modified Lorentzian. The fit
({\itshape dashed line}) is produced within a given range (1 $\leq$
$\ell$ $\leq$ 250). The function is extended throughout the full
data range for illustration purposes ({\itshape dotted line}). The
FWHM ({\itshape solid line}) is shown within the supergranule
feature.} \label{fig:sizes}
\end{center}
\end{figure}

Using the method described by \citet{hat00}, power spectra were
produced for each Dopplergram contained in the 1996 and 2008
datasets. Mean spectra were produced for both years by averaging
over all the individual spectra. A best-fit to the supergranule
feature (1 $\leq$ $\ell$ $\leq$ 250) was deduced, using a modified
Lorentzian function (Figure \ref{fig:sizes}). Although the fit
deviates from the data at higher wavenumbers, this is of no
consequence to the study as only the behavior near the peak is of
interest. The location of the peak in the fitting determines
$\ell_{max}$ and the FWHM of the supergranule distribution. From the
peak and FWHM, the typical diameter and size range of supergranules
was determined.

\section{Results} \label{Results}
The supergranule velocity analysis was performed out to an angular
distance of $\rho$~$\sim$~30 degrees. The 2008 data produced radial
and horizontal RMS velocity components of $v_r$ = 40 m/s and $v_h$ =
303 m/s, with $v_r$/$v_h$ = 0.13. The 1996 data produced $v_r$ = 24
and $v_h$ = 278 m/s, with $v_r$/$v_h$ = 0.09. These values have an
error of $\pm$ 1 m/s.

For 2008, the correlation analysis gave a mean decay coefficient of
0.056 $\pm$ 0.001 correlation coefficient units per hour.
Calculating decay-times from each coefficient, the harmonic mean
calculation gave a 1/$e$ decay time of 18 hours. This value was also
derived using the same process for the 1996 data.

For the supergranule size analysis, the 2008 dataset gave a peak
wavenumber of $\ell$ = 124 $\pm$ 1 and an FWHM of $\Delta\ell$ = 168
$\pm$ 5. These results respectively correspond to an average
supergranule diameter of 35.2 Mm and a diameter range of 18.7-67.2
Mm. For 1996, the peak wavenumber was found to be $\ell$ = 121 $\pm$
1 and an FWHM of $\Delta\ell$ = 160 $\pm$ 3, corresponding to an
average diameter of 36.1 Mm with a diameter range of 19.5-68.3 Mm.

\section{Analysis, Discussion and Conclusion} \label{sec:Analysis}

While the 1/$e$ lifetimes are found to be the same for both years,
the supergranule cell sizes tend to be slightly smaller during 2008.
A significant discrepancy, however, is found within the velocity
analysis.

Although the 1996 results are in line with those found by
\citet{hat02}, the 2008 velocity values are considerably greater.
Image defocusing during 1996 \citep{kor04} may influence the
velocity values contained within the Dopplergrams by altering the
resolution of the instrument. Active regions present on the disk may
contribute to spurious velocity values within the calculation
\citep{liu01}. Any images displaying significant activity are
currently removed from the analysis, whereas future analyses will
use magnetograms to mask out these regions so the images may be
included.

Comparing the past two solar minima, the decay rate and decay-times
of the supergranule pattern have not changed. Missing images within
a day's data currently means removing that day from the analysis,
which reduces the sampling for the averaging. Future work will
update the analysis process so that loss of data within a day will
not mean that the whole day needs to be skipped.

The size analysis is much less problematic although it can be
improved using a statistical analysis of the extracted parameters
for each Dopplergram-derived spectrum. This is planned for the
future along with producing similar statistics for the lifetimes and
velocity analyses.

\acknowledgements Peter Williams is supported by the Solar Dynamics
Observatory via the NASA Postdoctoral Program, managed by Oak Ridge
Associated Universities of Oak Ridge, Tennessee.  SOHO is a project
of international cooperation between ESA and NASA.

\end{document}